\documentclass[journal=jacsat,manuscript=article]{achemso}

\usepackage{graphicx}
\usepackage{amsmath}
\usepackage{bm}
\usepackage{amssymb}
\usepackage{units}
\usepackage{color}
\usepackage{setspace}

\author{S. Aeschlimann}
\affiliation{University of Regensburg, Institute of Experimental and Applied Physics, Regensburg, Germany}
\alsoaffiliation{Max Planck Institute for the Structure and Dynamics of Matter, Center for Free Electron Laser Science, Hamburg, Germany}

\author{S. A. Sato}
\affiliation{Max Planck Institute for the Structure and Dynamics of Matter, Center for Free Electron Laser Science, Hamburg, Germany}
\alsoaffiliation{Center for Computational Sciences, University of Tsukuba, Tsukuba 305-8577, Japan}

\author{R. Krause}
\affiliation{University of Regensburg, Institute of Experimental and Applied Physics, Regensburg, Germany}
\alsoaffiliation{Max Planck Institute for the Structure and Dynamics of Matter, Center for Free Electron Laser Science, Hamburg, Germany}

\author{M. Ch{\'a}vez-Cervantes}
\affiliation{Max Planck Institute for the Structure and Dynamics of Matter, Center for Free Electron Laser Science, Hamburg, Germany}

\author{U. De Giovannini}
\affiliation{Max Planck Institute for the Structure and Dynamics of Matter, Center for Free Electron Laser Science, Hamburg, Germany}
\alsoaffiliation{Nano-Bio Spectroscopy Group,  Departamento de Fisica de Materiales, Universidad del Pa{\'i}s Vasco UPV/EHU- 20018 San Sebasti{\'a}n, Spain}

\author{H. H\"ubener}
\affiliation{Max Planck Institute for the Structure and Dynamics of Matter, Center for Free Electron Laser Science, Hamburg, Germany}

\author{S. Forti}
\affiliation{Center for Nanotechnology Innovation @NEST, Istituto Italiano di Tecnologia, Piazza San Silvestro 12, 56127 Pisa, Italy}

\author{C. Coletti}
\affiliation{Center for Nanotechnology Innovation @NEST, Istituto Italiano di Tecnologia, Piazza San Silvestro 12, 56127 Pisa, Italy}
\alsoaffiliation{Graphene Labs, Istituto Italiano di Tecnologia, Via Morego 30, 16163 Genova, Italy}

\author{K. Hanff}
\affiliation{Max Planck Institute for the Structure and Dynamics of Matter, Center for Free Electron Laser Science, Hamburg, Germany}
\alsoaffiliation{Institut f\"ur Experimentelle und Angewandte Physik, Christian-Albrechts-Universit\"at zu Kiel, 24098 Kiel, Germany}

\author{K. Rossnagel}
\affiliation{Institut f\"ur Experimentelle und Angewandte Physik, Christian-Albrechts-Universit\"at zu Kiel, 24098 Kiel, Germany}
\alsoaffiliation{Ruprecht Haensel Laboratory, Deutsches Elektronen-Synchrotron DESY, 22607 Hamburg, Germany}

\author{A. Rubio}
\affiliation{Max Planck Institute for the Structure and Dynamics of Matter, Center for Free Electron Laser Science, Hamburg, Germany}
\alsoaffiliation{Nano-Bio Spectroscopy Group,  Departamento de Fisica de Materiales, Universidad del Pa{\'i}s Vasco UPV/EHU- 20018 San Sebasti{\'a}n, Spain}
\alsoaffiliation{Center for Computational Quantum Physics (CCQ), The Flatiron Institute, 162 Fifth avenue, New York NY 10010.}

\author{I. Gierz}
\email{isabella.gierz@ur.de}
\affiliation{University of Regensburg, Institute of Experimental and Applied Physics, Regensburg, Germany}

\title{On the survival of Floquet-Bloch states in the presence of scattering}

\abbreviations{tr-ARPES, TDDFT, Rabi model}
\keywords{Floquet-Bloch states, dissipation}

\begin{document}

\pagebreak

\begin{abstract}

Floquet theory has spawned many exciting possibilities for electronic structure control with light with enormous potential for future applications. The experimental realization in solids, however, largely remains pending. In particular, the influence of scattering on
the formation of Floquet-Bloch states remains poorly understood. Here we combine time- and angle-resolved photoemission spectroscopy with time-dependent density functional theory and a two-level model with relaxation to investigate the survival of Floquet-Bloch states in the presence of scattering. We find that Floquet-Bloch states will be destroyed if scattering --- activated by electronic excitations --- prevents the Bloch electrons from following the driving field coherently. The two-level model also shows that Floquet-Bloch states reappear at high field intensities where energy exchange with the driving field dominates over energy dissipation to the bath. Our results clearly indicate the importance of long scattering times combined with strong driving fields for the successful realization of various Floquet phenomena.
\end{abstract}

\pagebreak


With the recent development of strong-field Terahertz and mid-infrared laser sources, Floquet engineering, where the coherent interaction of strong light fields with Bloch electrons inside a solid is used to manipulate the band structure of the material, becomes a viable approach for non-equilibrium materials design. Floquet theory predicts a number of fascinating phenomena including dynamical localization of charge carriers \cite{DunlapPhysRevB1986} and light-induced topological phase transitions \cite{OkaPhysRevB2009,LindnerNatPhys2011}. Many of these phenomena have been observed in optical lattices \cite{LignierPhysRevLett2007,JotzuNature2014,EckardtRevModPhys2017}. In real solids, however, the ability of the Bloch electrons to coherently follow the driving field is limited by scattering. This has restricted the experimental observation of Floquet phenomena to materials with extremely long scattering times such as the topological insulator Bi$_2$Se$_3$ \cite{WangScience2013,MahmoodNatPhys2016} and exfoliated graphene at low temperatures \cite{McIverNatPhys2020}. The possible survival of Floquet phenomena in the presence of dissipation is a hotly debated topic \cite{DehghaniPRB2014} due to its relevance for future applications and the possibility to use dissipation to stabilize novel Floquet phases such as the time crystal \cite{WilczekPhysRevLett2012,ElsePhysRevLett2016,ZhangNature2017,ChoiNature2017} or topological phases \cite{DehghaniPRB2015}. 

Here, we use time- and angle-resolved photoemission spectroscopy (tr-ARPES) combined with time-dependent density functional theory (TDDFT) and a two-level model with relaxation to investigate the survival of Floquet-Bloch states in the presence of scattering. We find good agreement between TDDFT simulations and tr-ARPES measurements for sub-gap excitation of the bulk semiconductor WSe$_2$. In the case of graphene, however, TDDFT simulations predict the opening of various band gaps in the electronic dispersion that are not observed in the tr-ARPES measurements. We resolve this discrepancy with the help of a simple model where a resonantly driven two level system is coupled to a bath \cite{ArecchiIEEE1965,MaierBook2007,SatoPhysRevB2019,SatoJPhysB2020}. This model shows that Floquet-Bloch states will be destroyed if scattering --- activated by electronic excitations --- prevents the Bloch electrons from following the driving field coherently. The model also shows that Floquet-Bloch states reappear at high field intensities where energy exchange with the driving field dominates over energy dissipation to the bath.

The experimental observation of Floquet-Bloch states with tr-ARPES is impeded by the fact that in the presence of a strong driving laser field photoemission occurs from Floquet-Bloch states (photon-dressed initial states) to Volkov states (photon-dressed free-electron final states) \cite{MahmoodNatPhys2016, ParkPhysRevA2014, MadsenAmJPhys2005}. Photon-dressing of both the initial and the final state results in the formation of replica bands in the photoelectron spectrum that are separated from the original band structure by integer multiples of the drive photon energy. For the experimental geometry employed in the present study, the Volkov contribution to the first order replica bands is zero for s-polarized (sp) driving pulses (see Supporting Information). In this case, the experimental observation of first order replica bands with tr-ARPES provides direct evidence for the formation of Floquet-Bloch states.

\begin{figure}
	\center
		\includegraphics[width = 1\columnwidth]{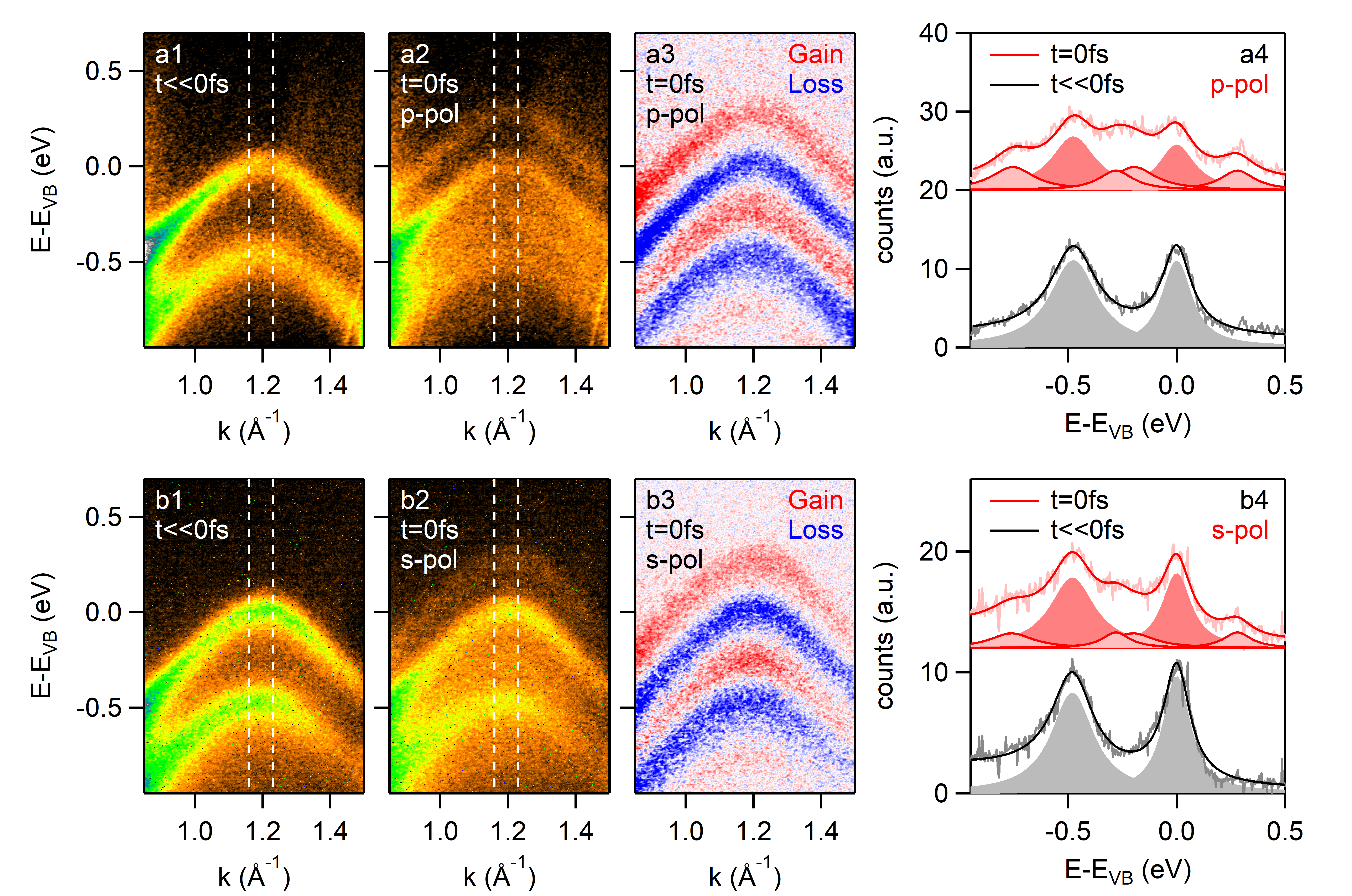}
  \caption{Rows a and b show the tr-ARPES data for WSe$_2$ for pp and sp driving pulses, respectively, at $\hbar\omega_{\text{drive}}=280$\,meV with a peak driving field of $E_{\text{vac}}=2.1$\,MV/cm. Columns 1 and 2 show the photocurrent at negative pump-probe delay and at t=0\,fs, respectively. The dashed white lines mark the momentum range for the energy distribution curves (EDCs) in Column 4. Column 3 shows the drive-induced changes of the photocurrent at $t=0$\,fs. These data were obtained by subtracting the data in Column 1 from the data in Column 2. Column 4 shows energy distribution curves (EDCs) extracted at the momentum range indicated by the dashed white lines in Columns 1 and 2 together with Lorentzian fits. Continuous light (dark) lines are the data (fit). Filled gray areas show the individual Lorentzians at negative delay. Filled light (dark) red areas show the individual Lorentzians of the sidebands (main bands).}
  \label{fig_WS2_exp}
\end{figure}

In Fig. \ref{fig_WS2_exp} we present tr-ARPES snapshots of the valence band of bulk WSe$_2$ measured along the $\Gamma$K-direction in the vicinity of the K-point of the Brillouin zone. Figures \ref{fig_WS2_exp}(a1) and (a2) show the measured band structure without and with p-polarized (pp) mid-infrared (MIR) driving field, respectively. The corresponding drive-induced changes of the photocurrent obtained by subtracting Fig. \ref{fig_WS2_exp}(a1) from Fig. \ref{fig_WS2_exp}(a2) are shown in Fig. \ref{fig_WS2_exp}(a3). Dashed white lines in Figs. \ref{fig_WS2_exp}(a1) and (a2) indicate the area of integration for the energy distribution curves (EDCs) presented in Fig. \ref{fig_WS2_exp}(a4). The corresponding data for sp driving pulses are presented in Figs. \ref{fig_WS2_exp}(b1)-(b4).

We observe clear indications for the formation of replica bands for both pp and sp driving pulses. As discussed in the Supporting Information, the observation of replica bands for sp driving pulses indicates the formation of Floquet-Bloch states, while the replica bands appearing for pp driving pulses contain contributions from both Floquet-Bloch and Volkov states. From Lorentzian fits to the EDCs in Fig. \ref{fig_WS2_exp}(b1)-(b4) we obtain replica band intensities of $(23.3\pm0.3)$\% and $(15.0\pm0.5)$\% for pp and sp driving pulses, respectively. The line width of the bands is found to be unaffected by the presence of the driving field within the error bars. Details about the data fitting are given in the Supporting Information.

\begin{figure}
	\center
		\includegraphics[width = 1\columnwidth]{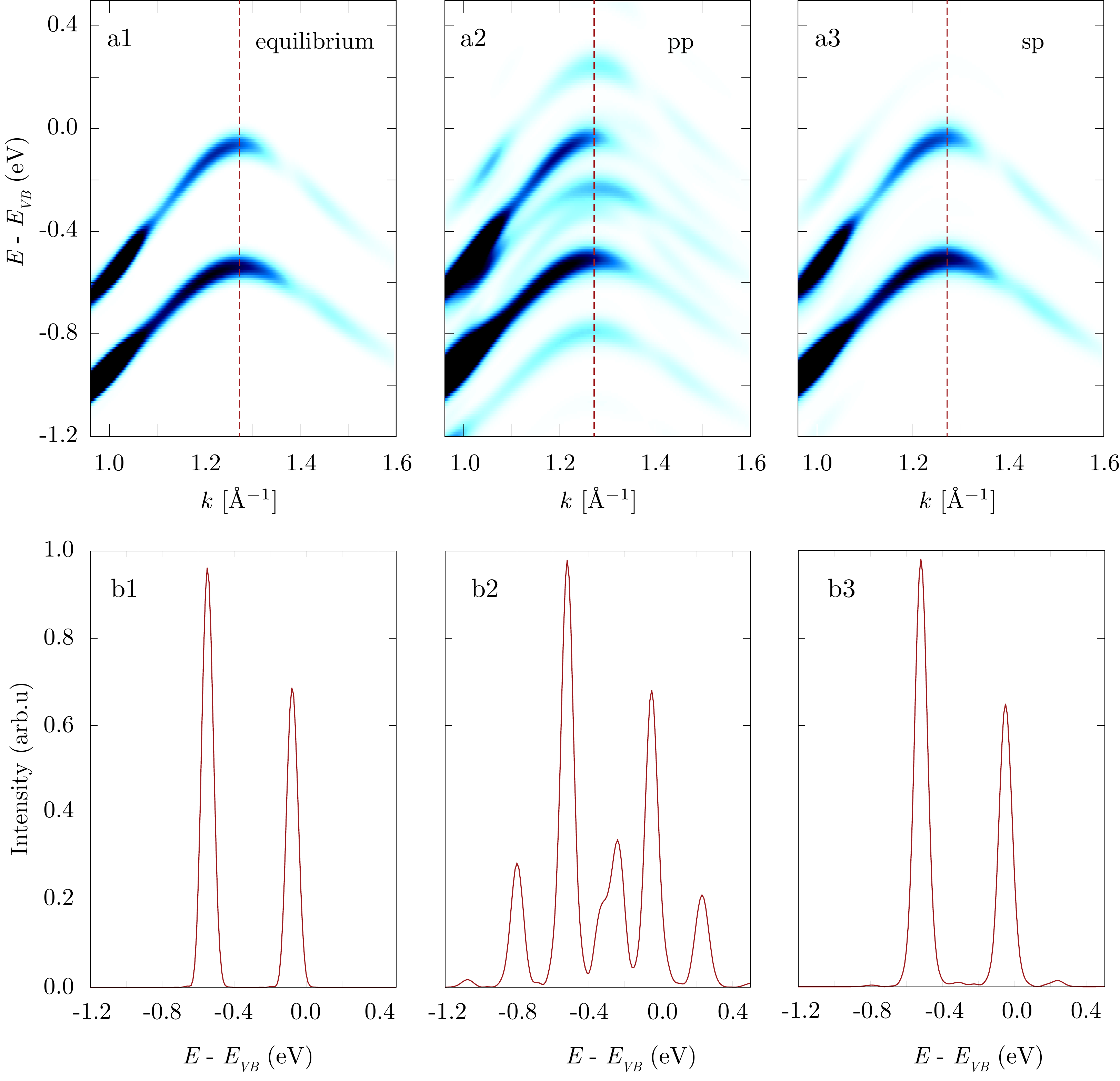}
  \caption{Simulated ARPES spectra for WSe$_2$ at equilibrium (a1), in the presence of a pp driving pulse (a2), and in the presence of a sp driving pulse (a3). Dashed red lines in a1-a3 indicate the positions for the energy distribution curves (EDCs) in panels b1-b3. EDC through the K-point at equilibrium (b1), in the presence of a pp driving pulse (b2), and in the presence of a sp driving pulse (b3). }
  \label{fig_WSe2_theory}
\end{figure}

In Fig. \ref{fig_WSe2_theory} we present TDDFT simulations for direct comparison with the tr-ARPES data in Fig. \ref{fig_WS2_exp}. Figure \ref{fig_WSe2_theory}(a1) shows the simulated spectrum without mid-infrared drive. Figures \ref{fig_WSe2_theory}(a2) and (a3) show the corresponding spectra in the presence of pp and sp driving pulses, respectively, with field strength and driving frequency matching the experimental values at the sample surface. In Figs. \ref{fig_WSe2_theory} (b1)-(b3) we show EDCs extracted along the dashed red lines in Figs. \ref{fig_WSe2_theory} (a1)-(a3). In good agreement with the tr-ARPES experiments in Fig. \ref{fig_WS2_exp} we find that, in the presence of both pp and sp driving pulses, replica bands appear in the simulated ARPES spectra. A comparison of the theoretical and experimental EDCs reveals that the theoretical replica band intensity is similar to the experimental value for pp driving pulses and significantly smaller than the experimental value for sp driving pulses. 

Having demonstrated the capability of our tr-ARPES setup to generate and resolve Floquet-Bloch states we now turn to the fascinating scenario of a light-induced topological phase transition in graphene. According to Ref. \cite{OkaPhysRevB2009} strong driving with circularly polarized (cp) light fields is predicted to open a gap at the Dirac point and turn graphene into a topological insulator. Additional band gaps (Rabi gaps in the following) are predicted to open away from the Dirac point where the unperturbed band structure crosses the $n$th order Floquet replica band \cite{SyzranovPhysRevB2008, OkaPhysRevB2009}. Because the matrix element for interband transitions in graphene is highly anisotropic \cite{TrushinEurophysLett2011, MalicPhysRevB2011, AeschlimannPhysRevB2017} with nodes (maxima) in the direction parallel (perpendicular) to the polarization of the driving field, these Rabi gaps are biggest along the direction in momentum space that is perpendicular to the polarization of the driving field. This is the case for sp driving pulses in the present study.

\begin{figure}
	\center
		\includegraphics[width = 1\columnwidth]{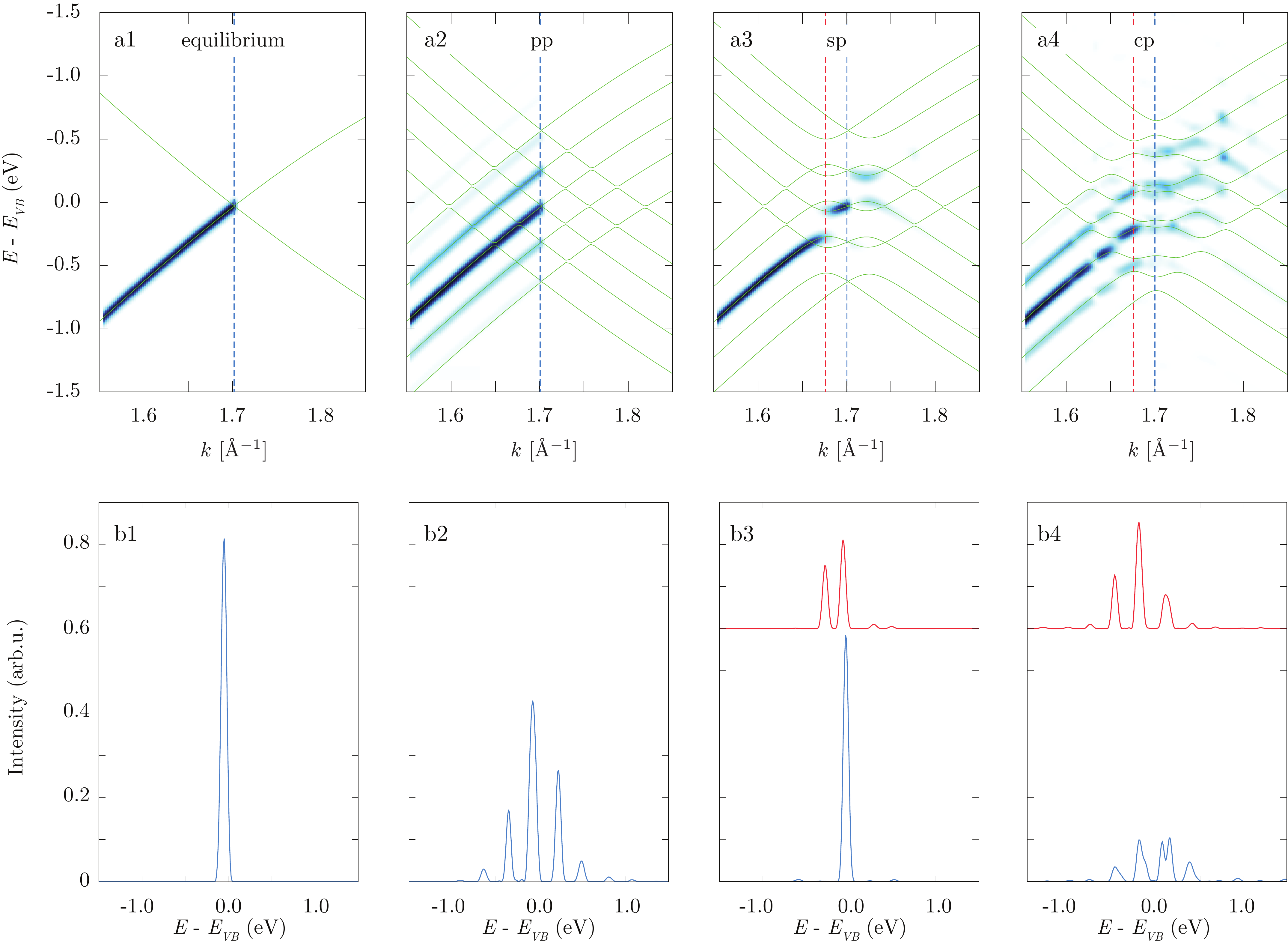}
  \caption{Simulated ARPES spectra for graphene together with initial state spectra (green lines) at equilibrium (a1), in the presence of a pp driving pulse (a2), in the presence of a sp driving pulse (a3), and in the presence of a cp driving pulse (a4). Blue and red dashed lines indicate the positions for the energy distribution curves (EDCs) in panels b1-b4. EDC through the Dirac point at equilibrium (b1), and in the presence of a pp driving pulse (b2). EDC through the Dirac point (blue) and at the position where the Rabi splitting is most pronounced (red) in the presence of a sp driving pulse (b3), and in the presence of a cp driving pulse (b4). }
  \label{fig_Graphene_theory}
\end{figure}

In Fig. \ref{fig_Graphene_theory} we present TDDFT simulations that illustrate how mid-infrared driving pulses with different polarizations are expected to affect the ARPES spectrum of graphene. Again, driving frequency as well as field strength were chosen to match the experimental conditions. Fig. \ref{fig_Graphene_theory}(a1) shows the equilibrium spectrum of graphene. Note that the right branch of the Dirac cone is invisible due to photoemission matrix element effects \cite{ShirleyPhysRevB1995, DaimonJElectronSpectroscRelatPhenom1995}. In the presence of a pp driving field the spectrum in Fig. \ref{fig_Graphene_theory}(a2) shows strong replica bands that, according to the model from \cite{ParkPhysRevA2014}, contain contributions from both Floquet-Bloch and Volkov states (see Supporting Information). For sp driving fields (Fig. \ref{fig_Graphene_theory}a3) the replica bands are found to be much weaker due to their pure Floquet-Bloch character. Also, the simulations clearly show the predicted Rabi gaps. The spectrum for the topologically non-trivial state in the presence of a circularly polarized (cp) driving pulse is shown in Fig. \ref{fig_Graphene_theory}(a4). Aside from replica bands and Rabi gaps the spectrum shows a pronounced band gap at the Dirac point. In order to get a better impression of the intensity of the replica bands as well as the size of the dynamical band gaps, Figs. \ref{fig_Graphene_theory}(b1)-(b4) present EDCs extracted along the dashed vertical lines in Figs. \ref{fig_Graphene_theory}(a1)-(a4). From Lorentzian fits to the EDC in Fig. \ref{fig_Graphene_theory}(b3) we extract a Rabi gap of 200\,meV for the sp drive. For the cp drive Lorentzian fits of the EDCs in Fig. \ref{fig_Graphene_theory}(b4) yield 100\,meV for the Rabi gap and 60\,meV for the gap at the Dirac point.

\begin{figure}
	\center
		\includegraphics[width = 1\columnwidth]{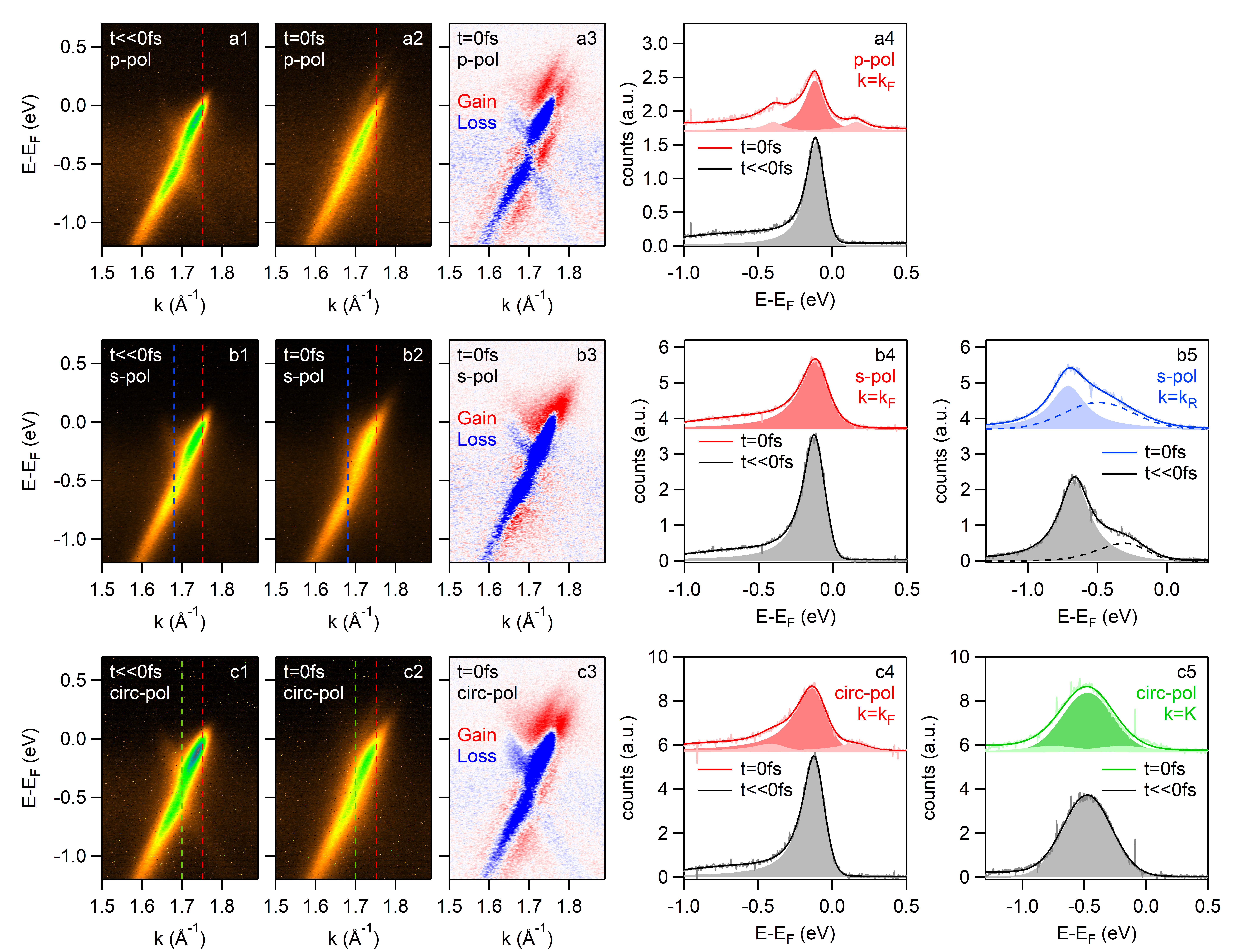}
  \caption{Rows a, b, and c show the tr-ARPES data for graphene for pp, sp, and cp driving pulses, respectively, at $\hbar\omega_{\text{drive}}=280$\,meV with a peak driving field of $E_{\text{vac}}=2.2$\,MV/cm. Columns 1 and 2 show the photocurrent at negative pump-probe delay and at t=0\,fs, respectively. Dashed lines mark the positions for the energy distribution curves (EDCs) in Columns 4 and 5. Column 3 shows the drive-induced changes of the photocurrent at t=0\,fs. These data were obtained by subtracting the data in Column 1 from the data in Column 2. Column 4 shows EDCs at the Fermi momenta from Column 1 and 2 together with Lorentzian fits. b5 shows EDCs through the Dirac point in b1 and b2 together with Gaussian fits. The dashed lines correspond to the second branch of the Dirac cone the intensity of which is suppressed due to photoemission matrix element effects. c5 shows EDCs at the momenta where the Rabi splitting is expected to occur in c1 and c2 together with Lorentzian fits. Filled gray areas show the individual peaks at negative delay. Filled light (dark) colored areas show the individual peaks of the sidebands (main bands).}
  \label{fig_Graphene_exp}
\end{figure}

In Fig. \ref{fig_Graphene_exp} we show the experimental data for direct comparison. The photocurrent at negative pump-probe delay is shown in column 1. In agreement with theory only one of the two branches of the Dirac cone is visible in the spectrum. In contrast to the simulations that were performed for a neutral graphene layer, the epitaxial graphene samples used for the tr-ARPES experiment are n-doped with the Dirac point 0.4\,eV below the Fermi level (see Supporting Information). The second column of Fig. \ref{fig_Graphene_exp} shows the photocurrent at zero pump-probe delay. The data for pp, sp, and cp driving pulses are shown in row a, b, and c, respectively. Column 3 shows the drive-induced changes of the photocurrent for the three light polarizations. 

To assess the formation of replica bands we extract EDCs close to the Fermi wave vector $k=k_F$ where the measured band structure is sharpest (column 4 of Fig. \ref{fig_Graphene_exp}). Replica bands are clearly resolved in Fig. \ref{fig_Graphene_exp}(a4) for pp driving pulses. The EDC for the sp drive, however, shows a single broad peak [Fig. \ref{fig_Graphene_exp}(b4)]. The EDC in Fig. \ref{fig_Graphene_exp}(c4) for cp driving pulses shows both broadening and replica bands, albeit not as well resolved as in Fig. \ref{fig_Graphene_exp}(a4). The EDC in Fig. \ref{fig_Graphene_exp}(b5) for sp driving was extracted at the position of the putative Rabi gap $k=k_R$. However, no gap is resolved in the data. The same holds for the topologically non-trivial gap for cp light: No gap is resolved in the EDC through the Dirac point at $k=K$ shown in Fig. \ref{fig_Graphene_exp}(c5). 

In order to extract the intensity of the replica bands as well as the peak widths the EDCs in Fig. \ref{fig_Graphene_exp} were fitted with an appropriate number of Lorentzians except the EDCs through the Dirac point that were fitted with a Gaussian. The fit results are summarized in table \ref{tab_FitResultsgraphene}. 

\begin{table}
	\centering
		\begin{tabular}{|l|c|c|c|}
		\hline
			&	FWHM & intensity of replica bands \\
		\hline
		\hline
		pp $t<0$\,fs $k=k_F$ & $153\pm2$\,meV &  0\,\% \\
		\hline
		pp $t=0$\,fs $k=k_F$ & $153\pm0$\,meV & $12.5\pm0.5$\,\% \\
		\hline
		\hline
		sp $t<0$\,fs $k=k_F$ & $189\pm2$\,meV & 0\,\% \\
		\hline
		sp $t=0$\,fs $k=k_F$ & $247\pm4$\,meV & 0\,\% \\
		\hline
		sp $t<0$\,fs $k=k_R$ & $269\pm6$\,meV & 0\,\% \\
		\hline
		sp $t=0$\,fs $k=k_R$ & $295\pm15$\,meV & 0\,\% \\
		\hline
		\hline
		cp $t<0$\,fs $k=k_F$ & $198\pm3$\,meV & 0\,\% \\
		\hline
		cp $t=0$\,fs $k=k_F$ & $296\pm5$\,meV & $9.3\pm0.3$\,\% \\
		\hline
		cp $t<0$\,fs $k=K$ & $466\pm12$\,meV & 0\,\% \\
		\hline
		cp $t=0$\,fs $k=K$ & $480\pm11$\,meV & $7\pm2$\,\% \\
		\hline
		\end{tabular}
	\caption{Fit results for the EDCs from Fig. \ref{fig_Graphene_exp}.}
	\label{tab_FitResultsgraphene}
\end{table}

Our findings can be summarized as follows: (1) We are able to resolve replica bands close to the Fermi edge whenever the driving field contains a pp component (i.e. pp and cp light). The intensity of the replica bands is found to be higher for pp than for cp light. (2) We observe a strong broadening of the spectra whenever the driving field contains a sp component (i.e. sp and cp light). This broadening is more pronounced for sp than for cp driving pulses. (3) We cannot resolve any band gap opening at $k=k_R$ or $k=K$. From the absence of Rabi gaps and replica bands for sp driving fields we conclude that no Floquet-Bloch states are formed in graphene under the present experimental conditions. Further, the absence of replica bands for sp driving fields suggests that the observed replica bands for pp driving fields likely originate from Volkov states alone.

Compared to the good agreement between theory and experiment in the case of WSe$_2$, the agreement is rather poor in the case of graphene. We attribute this to the fact that the present TDDFT simulations neglect scattering events which turn out to be crucial to understand the observed broadening of the Dirac cone in the presence of sp and cp driving pulses as discussed in detail below. For Floquet-Bloch states to form, the interaction between the Bloch electrons and the light field needs to be coherent. It has been suggested that coherence will be destroyed, if the scattering time of the Bloch electrons $\tau$ is shorter than or comparable to the period of the driving field $T_{\text{drive}}$ \cite{DunlapPhysRevB1986}. Considering this, our results suggest that coherent driving of the Bloch electrons is possible in WSe$_2$ but not in graphene.

One obvious difference between the two materials is that WSe$_2$ has a band gap of $\sim1$\,eV whereas graphene is a semimetal. At room temperature the WSe$_2$ valence band is completely filled and the conduction band is completely empty. Hence, the scattering phase space for Bloch electrons is zero and the scattering time is infinite. In neutral graphene the density of states at the Fermi level is zero. However, due to the absence of a band gap, any driving pulse with arbitrary frequency will generate electron-hole pairs. This increases the scattering phase space and the scattering time becomes finite. The situation is even worse in the case of epitaxial graphene that exhibits a strong n-doping with the Dirac point $\sim0.4$\,eV below the Fermi level resulting in a free carrier density of $n_e\approx10^{13}$\,cm$^{-2}$. Typical scattering times for photoexcited Dirac carriers in graphene are on the order of 10\,fs \cite{GierzPRL2015}, comparable to the period of the drive which is 15\,fs in the present study. Therefore, we conclude that scattering is detrimental for the formation of Floquet-Bloch states.

\begin{figure}
	\center
		\includegraphics[width = 1\columnwidth]{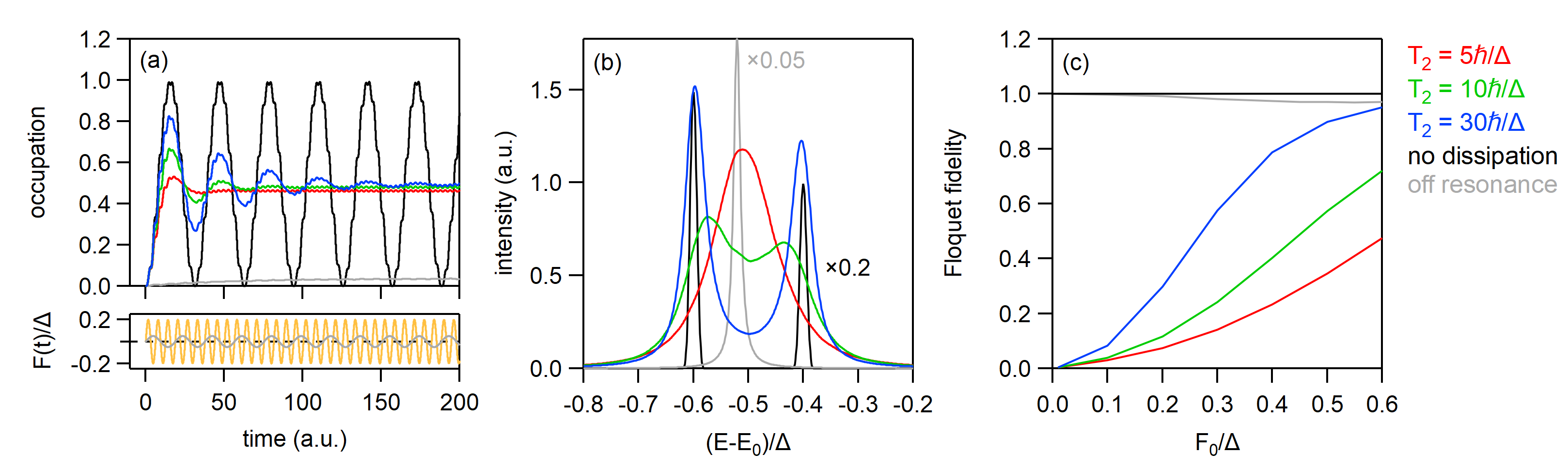}
  \caption{Theoretical results for a resonantly driven two-level system coupled to a bath. $T_1$ was set to 60\,$\hbar/\Delta$. The case without dissipation corresponds to $T_1=T_2=\infty$. (a) Population dynamics of the upper electronic level for different decoherence times $T_2$ (upper panel) in the presence of an external driving field (lower panel). (b) Quasienergy spectrum of the driven electronic system once a steady state is reached for different decoherence times $T_2$. Purple lines in (a) and (b) correspond to off-resonant driving with $F_0=0.05\Delta$. (c) Floquet fidelity of the non-equilibrium steady states as a function of field strength for different decoherence times $T_2$. }
  \label{fig_TwoLevelSystem}
\end{figure}

In the following we will present a simple model that allows us to investigate the influence of decoherence on Floquet-Bloch states and the corresponding quasienergy spectrum. We consider a resonantly driven two-level system with dissipation \cite{SatoJPhysB2020} the time propagation of which is described by the following quantum master equation

\begin{equation}
\frac{d}{dt}\rho(t)=\frac{[H(t),\rho(t)]}{i\hbar}+D[\rho(t)],
\label{equ_QME}
\end{equation}

where $\rho(t)$ is the density matrix of the system, $H(t)$ is the Hamiltonian, and $D[\rho(t)]$ is the relaxation operator. The Hamiltonian of the two-level system is given by

\begin{equation}
H(t)=\frac{\Delta}{2}\sigma_z+F_0\sin(\omega_{\text{drive}} t)\sigma_x,
\label{equ_H}
\end{equation}

where $\Delta$ is the energy gap of the two-level system, $\sigma_i$ are the Pauli matrices, and $F_0$ and $\omega_{\text{drive}}$ are the amplitude and the frequency of the driving field, respectively. We use a simple relaxation time approximation for the dissipation operator

\begin{equation}
D[\rho(t)]=\left(\begin{array}{rr}
-\frac{\rho_{ee}(t)}{T_1} & -\frac{\rho_{eg}(t)}{T_2}\\
-\frac{\rho_{ge}(t)}{T_2} & -\frac{\rho_{gg}(t)-1}{T_1}
\end{array} \right)
\label{equ_D}
\end{equation}

where $\rho_{ij}$ is a matrix element of the density matrix, where $i,j=g$ denotes the ground state and $i,j=e$ denotes the excited state. The longitudinal relaxation time $T_1$ accounts for the finite lifetime of the excited state, while the transverse relaxation time $T_2$ accounts for decoherence. Solving equation (\ref{equ_QME}) yields $\rho_{ee}(t)$ that is plotted in Fig. \ref{fig_TwoLevelSystem}a for different values of $T_2$ for $T_1=60\hbar/\Delta$. Without dissipation ($T_1=T_2=\infty$) $\rho_{ee}(t)$ is found to oscillate between zero and one with the Rabi frequency $\omega_R=F_0/\hbar$. In addition, we observe fast oscillations with the frequency of the driving field $\omega_{\text{drive}}$. For finite $T_1$ the Rabi oscillations are observed to be strongly damped with a lifetime that decreases with decreasing $T_2$. We also evaluate the quasienergy spectrum of the two-level system in the presence of dissipation as described in Ref. \cite{SatoJPhysB2020}. In Fig. \ref{fig_TwoLevelSystem}b we plot the quasienergie spectrum for different values of $T_2$ for $T_1=60\hbar/\Delta$. Without dissipation ($T_1=T_2=\infty$) we observe two sharp peaks that correspond to the Rabi splitting of the ground state. At finite $T_1$ the peaks are found to broaden and the Rabi splitting is found to decrease with decreasing $T_2$ until the two components merge into a single broad peak when $T_2<1/6$\,$T_R$ where $T_R=2\pi/\omega_R$.

In order to quantify to what degree the eigenstates of the driven dissipative system can be described by Floquet states we define the Floquet fidelity \cite{SatoPhysRevB2019} as $S_F=|\text{det}F|$ with the Floquet fidelity matrix $F$. The matrix elements of $F$

\begin{equation}
F_{ij}=\frac{1}{T}\int_0^T dt|\langle NO_i(t)|\Psi_{F,j}(t)\rangle|^2
\label{equ_FME}
\end{equation} 

are given by the absolute square of the overlap between natural orbitals $|NO_i(t)\rangle$ (eigenvectors of the single-particle density matrix) and the Floquet states $|\Psi_{F,j}(t)\rangle$ averaged over one period $T$ of the driving field. $S_F=1$ if all the natural orbitals are identical to Floquet states, and $S_F=0$ if natural orbitals and Floquet states do not overlap. In Fig. \ref{fig_TwoLevelSystem}c we plot the Floquet fidelity as a function of $T_2$ for different strengths of the driving field for $T_1=60\hbar/\Delta$. The Floquet fidelity is found to increase with increasing $T_2$ and with increasing field strength $F_0$. We find that, even in the presence of considerable decoherence, Floquet states can be recovered for sufficiently high driving fields. We also computed the situation where the driving frequency is much smaller than the gap (see red lines in Fig. \ref{fig_TwoLevelSystem}). In this case, the occupancy of the excited state remains zero (Fig. \ref{fig_TwoLevelSystem}a), the quasienergy spectrum shows a single sharp peak for the ground state (Fig. \ref{fig_TwoLevelSystem}b), and the Floquet fidelity is always one (Fig. \ref{fig_TwoLevelSystem}c).


We interpret our results as follows. As the valence band in semiconducting WSe$_2$ is completely filled the only possibility for the Bloch electrons to follow the driving field is a coherent motion through the Brillouin zone. In this case, Floquet-Bloch states are generated quite easily, in agreement with the observation of sharp replica bands for sp light in Fig. \ref{fig_WS2_exp}. In metallic graphene, however, a coherent motion of the driven Dirac carriers is only possible between two scattering events. If the scattering time $\tau$ becomes comparable to or shorter than the period of the drive $T_{\text{drive}}$, Floquet effects are expected to be smeared out by collisions \cite{DunlapPhysRevB1986}. Nevertheless, our model indicates that Floquet-Bloch states will survive in the presence of scattering, provided that the field strength is high enough. If the force that the driving field exerts on the electrons becomes stronger than the force due to a given scattering potential, the influence of the scattering potential will be negligible and Floquet states will be restored.

In the present study, the driving field strength was limited by the appearance of space charge effects in the tr-ARPES experiments. The maximum possible field strength turned out to be too low to generate Floquet-Bloch states. Alternatively, one might consider increasing the driving frequency such that, for a given scattering time, $\tau>T_{\text{drive}}$ \cite{DunlapPhysRevB1986}. However, the size of the Rabi gaps decreases as $1/\omega_{\text{drive}}$ \cite{SyzranovPhysRevB2008}, the size of the dynamical band gap at the Dirac point as $1/\omega_{\text{drive}}^3$ \cite{OkaPhysRevB2009}, and the intensity of the replica bands even as $1/\omega_{\text{drive}}^4$ \cite{ParkPhysRevA2014, MahmoodNatPhys2016, MadsenAmJPhys2005}. Therefore, it is unlikely that measurable Floquet effects survive at high driving frequencies.

Our findings seem to contradict the recent observation of a light-induced anomalous Hall effect in graphene using ultrafast transport experiments \cite{McIverNatPhys2020}. In contrast to our work on epitaxial graphene, the transport experiments were performed on exfoliated flakes where the Fermi level could be controlled via an applied gate voltage. The resulting carrier mobility was 10,000\,cm$^2$V$^{-1}$s$^{-1}$ in the vicinity of the Dirac point, one order of magnitude higher than typical carrier mobilities for graphene/SiC(0001) \cite{EmtsevNatMater2009}. Furthermore, the transport experiments were performed at 80\,K, whereas the tr-ARPES experiments were performed at room temperature. Due to these differences, it its quite likely that in the ultrafast transport experiments $\tau>T_{\text{drive}}$ while in the present tr-ARPES experiments $\tau<T_{\text{drive}}$.


In summary, we have used tr-ARPES to investigate the band structure changes induced by strong mid-infrared driving of WSe$_2$ and graphene. Good agreement between tr-ARPES and TDDFT simulations for WSe$_2$ indicates the formation of Floquet-Bloch states. In the case of graphene, however, TDDFT simulations predict the formation of replica bands, Rabi gaps, and a Rabi gap at the Dirac point, none of which are observed experimentally. Instead, tr-ARPES reveals a pronounced broadening of the spectral features that we attribute to decoherence via scattering supported by simulations based on a resonantly driven two-level system with dissipation.

Our results clearly reveal the practical limitations of Floquet engineering. Floquet-Bloch states are readily generated in semiconductors using subgap excitation. There, however, the induced band structure changes are trivial and leave the transport porperties unaffected. More intriguing phenomena such as light-induced topological phase transitions often rely on resonant driving of the material. These proposals can only be implemented successfully, if the scattering time of the respective excited states is long enough, limiting the approach to materials with extremely long scattering times. Topological insulators such as Bi$_2$Te$_3$ where scattering times in excess of 1\,ps have been reported \cite{ReimannNature2018} and exfoliated graphene at low temperature \cite{McIverNatPhys2020} might be among the few materials that fulfill the severe requirements of Floquet engineering.

\begin{acknowledgement}
This  work  was  supported  by  the  Deutsche Forschungsgemeinschaft (DFG, German Research Foundation) through CRC 925 (project 170620586), CRC 1277 (project 314695032), and the Cluster of Excellence 'CUI: Advanced Imaging of Matter'. Further, the work received funding from the European Research Council (starting grant 851280 and advanced grant 694097), the European Union Graphene Flagship under grant agreements nos. 785219 and 881603, Grupos Consolidados (IT1249-19), and JSPS KAKENHI (grant number JP20K14382). The Flatiron Institute is a division of the Simons Foundation.

\end{acknowledgement}

\begin{suppinfo}
The Supporting Information contains details about sample growth, tr-ARPES setup, data analysis, and simulations.
\end{suppinfo}



\clearpage
\pagebreak

\clearpage
\pagebreak

\section*{Supplementary Information}

\subsection*{Sample Growth}

Bulk 2H-WSe$_2$ samples were grown via standard chemical vapor transport reactions \cite{TravingPRB1997, BuckPRB2011}. For the growth of monolayer epitaxial graphene 6H-SiC wafers were graphitized in a modified Aixtron Black Magic reactor in Ar atmosphere at atmospheric pressure \cite{EmtsevNatMater2009} using the parameters reported in Ref. \cite{Rossi2DMater2016}. The resulting graphene samples were n-doped with the Fermi level $\sim$0.4\,eV above the Dirac point. After growth all samples were transported in air and inserted into ultra-high vacuum for the tr-ARPES experiments. In order to obtain a clean surface the WSe$_2$ samples were cleaved {\it in situ} using Kapton tape and the graphene samples were annealed at a temperature of 800$^\circ$C for 5\,min. 

\subsection*{Time- and Angle-Resolved Photoemission Spectroscopy (tr-ARPES)}

We used a commercial Titanium:Sapphire amplifier (Legend Elite Duo from Coherent) delivering 35\,fs pulses at a wavelength of 790\,nm with a repetition rate of 1\,kHz to generate femtosecond mid-infrared (MIR) pump and extreme ultraviolet (XUV) probe pulses. 10\,mJ of output power were sent into a commercial optical parametric amplifier (HE-TOPAS from Light Conversion) where the fundamental of the laser was converted into signal and idler pulses at 1.34\,$\mu$m and 1.92\,$\mu$m, respectively. These pulses were overlapped on a GaSe crystal for difference frequency generation (DFG) of the pump pulse at $\lambda_{\text{drive}}=4.45$\,$\mu$m ($\hbar\omega_{\text{drive}}=280$\,meV, $T_{\text{drive}}=15$\,fs). In order to block signal and idler pulses and to reduce the spectral width of the pump pulse the pump was sent through a bandpass filter centered at 4.45\,$\mu$m with a full width at half maximum (FWHM) of 200\,nm. The pump was then focused with a CaF$_2$ lens, yielding a pump fluence of $\sim2$\,mJ/cm$^2$ at the focus. The polarization of the pump pulse was controlled with a combination of $\lambda/2$ and $\lambda/4$ waveplates. 

XUV probe pulses were generated from the second harmonic of the remaining 2\,mJ of output power by high harmonics generation in Argon. The 7th harmonic at $\hbar\omega_{\text{probe}}=21.7$\,eV was selected with a time-preserving grating monochromator \cite{PolettoApplOptics2010} and focused onto the sample with a toroidal mirror. P-polarized XUV pulses were used to eject photoelectrons from the sample. The photocurrent as a function of kinetic energy and emission angle of the photoelectrons was measured with a hemispherical analyzer (Phoibos 100 from SPECS). The signal on the two-dimensional detector was converted into snapshots of the energy and momentum dependent spectral function $A(E,\vec{k_{\parallel}})$ of the material under investigation. The energy resolution was 150\,meV.

From the Gaussian full width at half maximum (FWHM) of the temporal profile of the replica band intensity we extract a pump-probe cross correlation of 300\,fs. With a nominal XUV probe pulse duration of 120\,fs we obtain a pump pulse duration of 270\,fs. Therefore, the experimental conditions satisfy $\text{FWHM}_{\text{drive}}>\text{FWHM}_{\text{probe}}>T_{\text{drive}}$, where $T_{\text{drive}}=15$\,fs is the period of the pump, required for the experimental observation of replica bands in periodically driven solids \cite{SentefNatCommun2014}.

From the experimental fluence of 2\,mJ/cm$^2$ and the driving pulse duration of 270\,fs we obtain a peak field strength of $E_{\text{vac}}=2.4$\,MV/cm at the focus in vacuum. The field strength in the surface of the sample was calculated via $E_{\text{sample}}=2E_{\text{vac}}/(1+\sqrt{\epsilon_{\infty}})$ where $\epsilon_{\infty}=6.7$ for SiC \cite{PatrickPhysRevB1970} and $\epsilon_{\infty}=15.6$ for WSe$_2$ \cite{Laturia2DMater2018}. Note that the dielectric constant for epitaxial graphene on SiC(0001) remains controversial with values ranging from $\epsilon_{\infty}=22$ \cite{WalterPRB2011} to $\epsilon_{\infty}=7.26$ \cite{HwangSciRep2012}. We would like to stress that using a different value of $\epsilon_{\infty}$ for our TDDFT simulations induces quantitative but no qualitative changes in the calculated ARPES spectra. Hence, the agreement with experiment cannot be improved by simply adapting the dielectric constant.

\subsection*{Data Analysis}

The experimental EDCs at negative pump-probe delay were fitted with the following function

\begin{equation}
f(E)=\left(\sum_{n}\text{Peak}(E,E_n,w_n)+\text{BG}_{\text{Shirley}}\right)f_{\text{FD}}(E,\mu,T)
\end{equation}

\noindent where Peak is either a Lorentzian or a Gaussian, $E_n$ and $w_n$ are peak position and full width at half maximum, respectively, BG$_{\text{Shirley}}$ is the Shirley background, f$_{\text{FD}}$ is the Fermi-Dirac distribution, $\mu$ is the chemical potential, and $T$ is the electronic temperature. The replica bands in the presence of the driving pulse were included like this

\begin{equation}
F=(1-2a)f(E)+af(E-\hbar\omega_{\text{drive}})+af(E+\hbar\omega_{\text{drive}})
\end{equation}

\noindent where $a$ is the intensity of the replica bands, and $\omega_{\text{drive}}$ is the driving frequency.

\subsection*{Time-Dependent Density Functional Theory (TDDFT)}

We compute the photoemission spectra using density-functional theory as implemented in the Octopus code \cite{TancogneJChemPhys2020}. In this method the time evolution of an electronic structure under the influence of pump and probe lasers in the dipole approximation is computed by propagating the Kohn-Sham equations in time \cite{KohnPhysRev1965, RungePRL1984}. The photoelectron spectrum is then obtained from the flux of ionised electronic states through an analyser surface placed at an appropriate distance away from the surface \cite{DeGiovanniniJChemTheoryComput2017}. The electronic structure of WSe$_2$ and graphene was obtained using the local density approximation for the DFT exchange-correlation potential and HGH pseudo potentials \cite{HartwigsenPRB1998}. The states of WSe$_2$ and graphene where represented in the unit cell on a real space grid with a spacing of 0.4\,Bohr and 0.36\,Bohr, respectively, and with k-point samplings of $6\times6\times1$ and $12\times12\times1$. In both cases the time evolution was evaluated with steps of 0.002\,fs and the analyser surface was placed at 90\,Bohr from the surface of the material. A complex absorbing boundary with a width of 30\,Bohr \cite{DeGiovanniniEurPhysJB2015} was used to prevent unphysical rescattering.

\subsection*{Photon-Dressed States}

Unfortunately, it is impossible to disentangle Volkov and Floquet-Bloch contributions to the replica band intensity of a photoelectron spectrum using TDDFT. Therefore, in order to calculate the intensity of the nth order Volkov replica band, we use the model from Madsen \cite{MadsenAmJPhys2005SI}. This model also includes the contribution of the ponderomotive energy $U_p$ which is neglected in the more recent model from Park et al. \cite{ParkPhysRevA2014SI}. In Madsen's model \cite{MadsenAmJPhys2005SI} the normalized intensity of the nth order Volkov replica band is given by

\begin{equation}
I_n/I_0\propto J_n^2\left(\vec{\alpha}\vec{q}_f, \frac{U_p}{2\omega_{\text{drive}}}\right)
\label{equ_InM}
\end{equation}

\noindent where $\vec{\alpha}=\vec{A}/\omega_{\text{drive}}$, $\vec{q}_f$ is the momentum of the photoelectron, and $U_p=A^2/4$ is the ponderomotive energy. The generalized Bessel functions $J_n(u,v)$ have the following property:

\begin{equation}
J_n(u,v)=\sum_{k=-\infty}^{\infty}J_{n-2k}(u)J_k(v).
\label{equ_Bessel}
\end{equation}

\noindent Also, $J_n(0)=0$ for $n\neq0$ and $J_0(0)=1$. With $u=\vec{\alpha}\vec{q}_f=0$ (which is the case for sp driving pulses in the present study) and $v=\frac{U_p}{2\omega}$ we get

\begin{align}
J_n(0,v)&=\sum_{k=-\infty}^{\infty}J_{n-2k}(0)J_k(v)\\
J_{n=2k}(0,v)&=\sum_{k=-\infty}^{\infty}J_{0}(0)J_k(v)\\
&=\sum_{k=-\infty}^{\infty}J_k(v)
\label{equ_InM2}
\end{align}

\noindent where we used that $J_{n-2k}(0)\neq0$ only if $n=2k$. This means that, if $\vec{\alpha}\vec{q}_f=0$, we will only get even-order Volkov replica bands. Hence, the Volkov contribution to the first order replica band is zero. Therefore, any first-order replica band that we observe in tr-ARPES using sp driving pulses necessarily needs to be a Floquet-Bloch state.

\bibliography{literature}

\providecommand{\latin}[1]{#1}
\makeatletter
\providecommand{\doi}
  {\begingroup\let\do\@makeother\dospecials
  \catcode`\{=1 \catcode`\}=2 \doi@aux}
\providecommand{\doi@aux}[1]{\endgroup\texttt{#1}}
\makeatother
\providecommand*\mcitethebibliography{\thebibliography}
\csname @ifundefined\endcsname{endmcitethebibliography}
  {\let\endmcitethebibliography\endthebibliography}{}
\begin{mcitethebibliography}{48}
\providecommand*\natexlab[1]{#1}
\providecommand*\mciteSetBstSublistMode[1]{}
\providecommand*\mciteSetBstMaxWidthForm[2]{}
\providecommand*\mciteBstWouldAddEndPuncttrue
  {\def\EndOfBibitem{\unskip.}}
\providecommand*\mciteBstWouldAddEndPunctfalse
  {\let\EndOfBibitem\relax}
\providecommand*\mciteSetBstMidEndSepPunct[3]{}
\providecommand*\mciteSetBstSublistLabelBeginEnd[3]{}
\providecommand*\EndOfBibitem{}
\mciteSetBstSublistMode{f}
\mciteSetBstMaxWidthForm{subitem}{(\alph{mcitesubitemcount})}
\mciteSetBstSublistLabelBeginEnd
  {\mcitemaxwidthsubitemform\space}
  {\relax}
  {\relax}

\bibitem[Dunlap and Kenkre(1986)Dunlap, and Kenkre]{DunlapPhysRevB1986}
Dunlap,~D.~H.; Kenkre,~V.~M. Dynamic localization of a charged particle moving
  under the influence of an electric field. \emph{Phys. Rev. B} \textbf{1986},
  \emph{34}, 3625--3633\relax
\mciteBstWouldAddEndPuncttrue
\mciteSetBstMidEndSepPunct{\mcitedefaultmidpunct}
{\mcitedefaultendpunct}{\mcitedefaultseppunct}\relax
\EndOfBibitem
\bibitem[Oka and Aoki(2009)Oka, and Aoki]{OkaPhysRevB2009}
Oka,~T.; Aoki,~H. Photovoltaic Hall effect in graphene. \emph{Phys. Rev. B}
  \textbf{2009}, \emph{79}, 081406\relax
\mciteBstWouldAddEndPuncttrue
\mciteSetBstMidEndSepPunct{\mcitedefaultmidpunct}
{\mcitedefaultendpunct}{\mcitedefaultseppunct}\relax
\EndOfBibitem
\bibitem[Lindner \latin{et~al.}(2011)Lindner, Refael, and
  Galitski]{LindnerNatPhys2011}
Lindner,~N.~H.; Refael,~G.; Galitski,~V. Floquet topological insulator in
  semiconductor quantum wells. \emph{Nat. Phys.} \textbf{2011}, \emph{7},
  490\relax
\mciteBstWouldAddEndPuncttrue
\mciteSetBstMidEndSepPunct{\mcitedefaultmidpunct}
{\mcitedefaultendpunct}{\mcitedefaultseppunct}\relax
\EndOfBibitem
\bibitem[Lignier \latin{et~al.}(2007)Lignier, Sias, Ciampini, Singh, Zenesini,
  Morsch, and Arimondo]{LignierPhysRevLett2007}
Lignier,~H.; Sias,~C.; Ciampini,~D.; Singh,~Y.; Zenesini,~A.; Morsch,~O.;
  Arimondo,~E. Dynamical Control of Matter-Wave Tunneling in Periodic
  Potentials. \emph{Phys. Rev. Lett.} \textbf{2007}, \emph{99}, 220403\relax
\mciteBstWouldAddEndPuncttrue
\mciteSetBstMidEndSepPunct{\mcitedefaultmidpunct}
{\mcitedefaultendpunct}{\mcitedefaultseppunct}\relax
\EndOfBibitem
\bibitem[Jotzu \latin{et~al.}(2014)Jotzu, Messer, Desbuquois, Lebrat,
  Uehlinger, Greif, and Esslinger]{JotzuNature2014}
Jotzu,~G.; Messer,~M.; Desbuquois,~R.; Lebrat,~M.; Uehlinger,~T.; Greif,~D.;
  Esslinger,~T. Experimental realization of the topological Haldane model with
  ultracold fermions. \emph{Nature} \textbf{2014}, \emph{515}, 237\relax
\mciteBstWouldAddEndPuncttrue
\mciteSetBstMidEndSepPunct{\mcitedefaultmidpunct}
{\mcitedefaultendpunct}{\mcitedefaultseppunct}\relax
\EndOfBibitem
\bibitem[Eckardt(2017)]{EckardtRevModPhys2017}
Eckardt,~A. Colloquium: Atomic quantum gases in periodically driven optical
  lattices. \emph{Rev. Mod. Phys.} \textbf{2017}, \emph{89}, 011004\relax
\mciteBstWouldAddEndPuncttrue
\mciteSetBstMidEndSepPunct{\mcitedefaultmidpunct}
{\mcitedefaultendpunct}{\mcitedefaultseppunct}\relax
\EndOfBibitem
\bibitem[Wang \latin{et~al.}(2013)Wang, Steinberg, Jarillo-Herrero, and
  Gedik]{WangScience2013}
Wang,~Y.~H.; Steinberg,~H.; Jarillo-Herrero,~P.; Gedik,~N. Observation of
  Floquet-Bloch States on the Surface of a Topological Insulator.
  \emph{Science} \textbf{2013}, \emph{342}, 453--457\relax
\mciteBstWouldAddEndPuncttrue
\mciteSetBstMidEndSepPunct{\mcitedefaultmidpunct}
{\mcitedefaultendpunct}{\mcitedefaultseppunct}\relax
\EndOfBibitem
\bibitem[Mahmood \latin{et~al.}(2016)Mahmood, Chan, Alpichshev, Gardner, Lee,
  Lee, and Gedik]{MahmoodNatPhys2016}
Mahmood,~F.; Chan,~C.-K.; Alpichshev,~Z.; Gardner,~D.; Lee,~Y.; Lee,~P.~A.;
  Gedik,~N. Selective scattering between Floquet-Bloch and Volkov states in a
  topological insulator. \emph{Nat. Phys.} \textbf{2016}, \emph{12}, 306\relax
\mciteBstWouldAddEndPuncttrue
\mciteSetBstMidEndSepPunct{\mcitedefaultmidpunct}
{\mcitedefaultendpunct}{\mcitedefaultseppunct}\relax
\EndOfBibitem
\bibitem[McIver \latin{et~al.}(2020)McIver, Schulte, Stein, Matsuyama, Jotzu,
  Meier, and Cavalleri]{McIverNatPhys2020}
McIver,~J.~W.; Schulte,~B.; Stein,~F.-U.; Matsuyama,~T.; Jotzu,~G.; Meier,~G.;
  Cavalleri,~A. Light-induced anomalous Hall effect in graphene. \emph{Nat.
  Phys.} \textbf{2020}, \emph{16}, 38\relax
\mciteBstWouldAddEndPuncttrue
\mciteSetBstMidEndSepPunct{\mcitedefaultmidpunct}
{\mcitedefaultendpunct}{\mcitedefaultseppunct}\relax
\EndOfBibitem
\bibitem[Dehghani \latin{et~al.}(2014)Dehghani, Oka, and
  Mitra]{DehghaniPRB2014}
Dehghani,~H.; Oka,~T.; Mitra,~A. Dissipative Floquet topological systems.
  \emph{Phys. Rev. B} \textbf{2014}, \emph{90}, 195429\relax
\mciteBstWouldAddEndPuncttrue
\mciteSetBstMidEndSepPunct{\mcitedefaultmidpunct}
{\mcitedefaultendpunct}{\mcitedefaultseppunct}\relax
\EndOfBibitem
\bibitem[Wilczek(2012)]{WilczekPhysRevLett2012}
Wilczek,~F. Quantum Time Crystals. \emph{Phys. Rev. Lett.} \textbf{2012},
  \emph{109}, 160401\relax
\mciteBstWouldAddEndPuncttrue
\mciteSetBstMidEndSepPunct{\mcitedefaultmidpunct}
{\mcitedefaultendpunct}{\mcitedefaultseppunct}\relax
\EndOfBibitem
\bibitem[Else \latin{et~al.}(2016)Else, Bauer, and Nayak]{ElsePhysRevLett2016}
Else,~D.~V.; Bauer,~B.; Nayak,~C. Floquet Time Crystals. \emph{Phys. Rev.
  Lett.} \textbf{2016}, \emph{117}, 090402\relax
\mciteBstWouldAddEndPuncttrue
\mciteSetBstMidEndSepPunct{\mcitedefaultmidpunct}
{\mcitedefaultendpunct}{\mcitedefaultseppunct}\relax
\EndOfBibitem
\bibitem[Zhang \latin{et~al.}(2017)Zhang, Hess, Kyprianidis, Becker, Lee,
  Smith, Pagano, Potirniche, Potter, Vishwanath, Yao, and
  Monroe]{ZhangNature2017}
Zhang,~J.; Hess,~P.~W.; Kyprianidis,~A.; Becker,~P.; Lee,~A.; Smith,~J.;
  Pagano,~G.; Potirniche,~I.-D.; Potter,~A.~C.; Vishwanath,~A.; Yao,~N.~Y.;
  Monroe,~C. Observation of a discrete time crystal. \emph{Nature}
  \textbf{2017}, \emph{543}, 217\relax
\mciteBstWouldAddEndPuncttrue
\mciteSetBstMidEndSepPunct{\mcitedefaultmidpunct}
{\mcitedefaultendpunct}{\mcitedefaultseppunct}\relax
\EndOfBibitem
\bibitem[Choi \latin{et~al.}(2017)Choi, Choi, Landig, Kucsko, Zhou, Isoya,
  Jelezko, Onoda, Sumiya, Khemani, von Keyserlingk, Yao, Demler, and
  Lukin]{ChoiNature2017}
Choi,~S.; Choi,~J.; Landig,~R.; Kucsko,~G.; Zhou,~H.; Isoya,~J.; Jelezko,~F.;
  Onoda,~S.; Sumiya,~H.; Khemani,~V.; von Keyserlingk,~C.; Yao,~N.~Y.;
  Demler,~E.; Lukin,~M.~D. Observation of discrete time-crystalline order in a
  disordered dipolar many-body system. \emph{Nature} \textbf{2017}, \emph{543},
  221\relax
\mciteBstWouldAddEndPuncttrue
\mciteSetBstMidEndSepPunct{\mcitedefaultmidpunct}
{\mcitedefaultendpunct}{\mcitedefaultseppunct}\relax
\EndOfBibitem
\bibitem[Dehghani \latin{et~al.}(2015)Dehghani, Oka, and
  Mitra]{DehghaniPRB2015}
Dehghani,~H.; Oka,~T.; Mitra,~A. Out-of-equilibrium electrons and the Hall
  conductance of a Floquet topological insulator. \emph{Phys. Rev. B}
  \textbf{2015}, \emph{91}, 155422\relax
\mciteBstWouldAddEndPuncttrue
\mciteSetBstMidEndSepPunct{\mcitedefaultmidpunct}
{\mcitedefaultendpunct}{\mcitedefaultseppunct}\relax
\EndOfBibitem
\bibitem[{Arecchi} and {Bonifacio}(1965){Arecchi}, and
  {Bonifacio}]{ArecchiIEEE1965}
{Arecchi},~F.; {Bonifacio},~R. Theory of optical maser amplifiers. \emph{IEEE
  Journal of Quantum Electronics} \textbf{1965}, \emph{1}, 169--178\relax
\mciteBstWouldAddEndPuncttrue
\mciteSetBstMidEndSepPunct{\mcitedefaultmidpunct}
{\mcitedefaultendpunct}{\mcitedefaultseppunct}\relax
\EndOfBibitem
\bibitem[Meier \latin{et~al.}(2007)Meier, Thomas, and Koch]{MaierBook2007}
Meier,~T.; Thomas,~P.; Koch,~S.~W. Coherent Semiconductor Optics.
  \emph{Springer} \textbf{2007}, \relax
\mciteBstWouldAddEndPunctfalse
\mciteSetBstMidEndSepPunct{\mcitedefaultmidpunct}
{}{\mcitedefaultseppunct}\relax
\EndOfBibitem
\bibitem[Sato \latin{et~al.}(2019)Sato, McIver, Nuske, Tang, Jotzu, Schulte,
  H\"ubener, De~Giovannini, Mathey, Sentef, Cavalleri, and
  Rubio]{SatoPhysRevB2019}
Sato,~S.~A.; McIver,~J.~W.; Nuske,~M.; Tang,~P.; Jotzu,~G.; Schulte,~B.;
  H\"ubener,~H.; De~Giovannini,~U.; Mathey,~L.; Sentef,~M.~A.; Cavalleri,~A.;
  Rubio,~A. Microscopic theory for the light-induced anomalous Hall effect in
  graphene. \emph{Phys. Rev. B} \textbf{2019}, \emph{99}, 214302\relax
\mciteBstWouldAddEndPuncttrue
\mciteSetBstMidEndSepPunct{\mcitedefaultmidpunct}
{\mcitedefaultendpunct}{\mcitedefaultseppunct}\relax
\EndOfBibitem
\bibitem[Sato \latin{et~al.}(2020)Sato, De~Giovannini, Aeschlimann, Gierz,
  H.\"ubener, and Rubio]{SatoJPhysB2020}
Sato,~S.~A.; De~Giovannini,~U.; Aeschlimann,~S.; Gierz,~I.; H.\"ubener,~H.;
  Rubio,~A. Floquet states in dissipative open quantum systems. \emph{J. Phys.
  B: At. Mol. Opt. Phys.} \textbf{2020}, \emph{53}, 225601\relax
\mciteBstWouldAddEndPuncttrue
\mciteSetBstMidEndSepPunct{\mcitedefaultmidpunct}
{\mcitedefaultendpunct}{\mcitedefaultseppunct}\relax
\EndOfBibitem
\bibitem[Park(2014)]{ParkPhysRevA2014}
Park,~S.~T. Interference in Floquet-Volkov transitions. \emph{Phys. Rev. A}
  \textbf{2014}, \emph{90}, 013420\relax
\mciteBstWouldAddEndPuncttrue
\mciteSetBstMidEndSepPunct{\mcitedefaultmidpunct}
{\mcitedefaultendpunct}{\mcitedefaultseppunct}\relax
\EndOfBibitem
\bibitem[Madsen(2005)]{MadsenAmJPhys2005}
Madsen,~L.~B. Strong-field approximation in laser-assisted dynamics.
  \emph{American Journal of Physics} \textbf{2005}, \emph{73}, 57--62\relax
\mciteBstWouldAddEndPuncttrue
\mciteSetBstMidEndSepPunct{\mcitedefaultmidpunct}
{\mcitedefaultendpunct}{\mcitedefaultseppunct}\relax
\EndOfBibitem
\bibitem[Syzranov \latin{et~al.}(2008)Syzranov, Fistul, and
  Efetov]{SyzranovPhysRevB2008}
Syzranov,~S.~V.; Fistul,~M.~V.; Efetov,~K.~B. Effect of radiation on transport
  in graphene. \emph{Phys. Rev. B} \textbf{2008}, \emph{78}, 045407\relax
\mciteBstWouldAddEndPuncttrue
\mciteSetBstMidEndSepPunct{\mcitedefaultmidpunct}
{\mcitedefaultendpunct}{\mcitedefaultseppunct}\relax
\EndOfBibitem
\bibitem[Trushin and Schliemann(2011)Trushin, and
  Schliemann]{TrushinEurophysLett2011}
Trushin,~M.; Schliemann,~J. Anisotropic photoconductivity in graphene.
  \emph{{EPL} (Europhysics Letters)} \textbf{2011}, \emph{96}, 37006\relax
\mciteBstWouldAddEndPuncttrue
\mciteSetBstMidEndSepPunct{\mcitedefaultmidpunct}
{\mcitedefaultendpunct}{\mcitedefaultseppunct}\relax
\EndOfBibitem
\bibitem[Malic \latin{et~al.}(2011)Malic, Winzer, Bobkin, and
  Knorr]{MalicPhysRevB2011}
Malic,~E.; Winzer,~T.; Bobkin,~E.; Knorr,~A. Microscopic theory of absorption
  and ultrafast many-particle kinetics in graphene. \emph{Phys. Rev. B}
  \textbf{2011}, \emph{84}, 205406\relax
\mciteBstWouldAddEndPuncttrue
\mciteSetBstMidEndSepPunct{\mcitedefaultmidpunct}
{\mcitedefaultendpunct}{\mcitedefaultseppunct}\relax
\EndOfBibitem
\bibitem[Aeschlimann \latin{et~al.}(2017)Aeschlimann, Krause,
  Ch\'avez-Cervantes, Bromberger, Jago, Mali\ifmmode~\acute{c}\else \'{c}\fi{},
  Al-Temimy, Coletti, Cavalleri, and Gierz]{AeschlimannPhysRevB2017}
Aeschlimann,~S.; Krause,~R.; Ch\'avez-Cervantes,~M.; Bromberger,~H.; Jago,~R.;
  Mali\ifmmode~\acute{c}\else \'{c}\fi{},~E.; Al-Temimy,~A.; Coletti,~C.;
  Cavalleri,~A.; Gierz,~I. Ultrafast momentum imaging of pseudospin-flip
  excitations in graphene. \emph{Phys. Rev. B} \textbf{2017}, \emph{96},
  020301\relax
\mciteBstWouldAddEndPuncttrue
\mciteSetBstMidEndSepPunct{\mcitedefaultmidpunct}
{\mcitedefaultendpunct}{\mcitedefaultseppunct}\relax
\EndOfBibitem
\bibitem[Shirley \latin{et~al.}(1995)Shirley, Terminello, Santoni, and
  Himpsel]{ShirleyPhysRevB1995}
Shirley,~E.~L.; Terminello,~L.~J.; Santoni,~A.; Himpsel,~F.~J.
  Brillouin-zone-selection effects in graphite photoelectron angular
  distributions. \emph{Phys. Rev. B} \textbf{1995}, \emph{51},
  13614--13622\relax
\mciteBstWouldAddEndPuncttrue
\mciteSetBstMidEndSepPunct{\mcitedefaultmidpunct}
{\mcitedefaultendpunct}{\mcitedefaultseppunct}\relax
\EndOfBibitem
\bibitem[Daimon \latin{et~al.}(1995)Daimon, Nakatani, Imada, and
  Suga]{DaimonJElectronSpectroscRelatPhenom1995}
Daimon,~H.; Nakatani,~T.; Imada,~S.; Suga,~S. Circular dichroism from
  non-chiral and non-magnetic materials observed with display-type spherical
  mirror analyzer. \emph{Journal of Electron Spectroscopy and Related
  Phenomena} \textbf{1995}, \emph{76}, 55--62, Proceedings of the Sixth
  International Conference on Electron Spectroscopy\relax
\mciteBstWouldAddEndPuncttrue
\mciteSetBstMidEndSepPunct{\mcitedefaultmidpunct}
{\mcitedefaultendpunct}{\mcitedefaultseppunct}\relax
\EndOfBibitem
\bibitem[Gierz \latin{et~al.}(2015)Gierz, Calegari, Aeschlimann,
  Ch\'avez~Cervantes, Cacho, Chapman, Springate, Link, Starke, Ast, and
  Cavalleri]{GierzPRL2015}
Gierz,~I.; Calegari,~F.; Aeschlimann,~S.; Ch\'avez~Cervantes,~M.; Cacho,~C.;
  Chapman,~R.~T.; Springate,~E.; Link,~S.; Starke,~U.; Ast,~C.~R.;
  Cavalleri,~A. Tracking Primary Thermalization Events in Graphene with
  Photoemission at Extreme Time Scales. \emph{Phys. Rev. Lett.} \textbf{2015},
  \emph{115}, 086803\relax
\mciteBstWouldAddEndPuncttrue
\mciteSetBstMidEndSepPunct{\mcitedefaultmidpunct}
{\mcitedefaultendpunct}{\mcitedefaultseppunct}\relax
\EndOfBibitem
\bibitem[Emtsev \latin{et~al.}(2009)Emtsev, Bostwick, Horn, Jobst, Kellogg,
  Ley, McChesney, Ohta, Reshanov, R\"ohrl, Rotenberg, Schmid, Waldmann, Weber,
  and Seyller]{EmtsevNatMater2009}
Emtsev,~K.~V.; Bostwick,~A.; Horn,~K.; Jobst,~J.; Kellogg,~G.~L.; Ley,~L.;
  McChesney,~J.~L.; Ohta,~T.; Reshanov,~S.~A.; R\"ohrl,~J.; Rotenberg,~E.;
  Schmid,~A.~K.; Waldmann,~D.; Weber,~H.~B.; Seyller,~T. Towards wafer-size
  graphene layers by atmospheric pressure graphitization of silicon carbide.
  \emph{Nature Materials} \textbf{2009}, \emph{8}, 203\relax
\mciteBstWouldAddEndPuncttrue
\mciteSetBstMidEndSepPunct{\mcitedefaultmidpunct}
{\mcitedefaultendpunct}{\mcitedefaultseppunct}\relax
\EndOfBibitem
\bibitem[Reimann \latin{et~al.}(2018)Reimann, Schlauderer, Schmid, Langer,
  Baierl, Kokh, Tereshchenko, Kimura, Lange, G\"udde, H\"ofer, and
  Huber]{ReimannNature2018}
Reimann,~J.; Schlauderer,~S.; Schmid,~C.~P.; Langer,~F.; Baierl,~S.;
  Kokh,~K.~A.; Tereshchenko,~O.~E.; Kimura,~A.; Lange,~C.; G\"udde,~J.;
  H\"ofer,~U.; Huber,~R. Subcycle observation of lightwave-driven Dirac
  currents in a topological surface band. \emph{Nature} \textbf{2018},
  \emph{562}, 396\relax
\mciteBstWouldAddEndPuncttrue
\mciteSetBstMidEndSepPunct{\mcitedefaultmidpunct}
{\mcitedefaultendpunct}{\mcitedefaultseppunct}\relax
\EndOfBibitem
\bibitem[Traving \latin{et~al.}(1997)Traving, Boehme, Kipp, Skibowski,
  Starrost, Krasovskii, Perlov, and Schattke]{TravingPRB1997}
Traving,~M.; Boehme,~M.; Kipp,~L.; Skibowski,~M.; Starrost,~F.;
  Krasovskii,~E.~E.; Perlov,~A.; Schattke,~W. Electronic structure of
  ${\mathrm{WSe}}_{2}$: A combined photoemission and inverse photoemission
  study. \emph{Phys. Rev. B} \textbf{1997}, \emph{55}, 10392--10399\relax
\mciteBstWouldAddEndPuncttrue
\mciteSetBstMidEndSepPunct{\mcitedefaultmidpunct}
{\mcitedefaultendpunct}{\mcitedefaultseppunct}\relax
\EndOfBibitem
\bibitem[Buck \latin{et~al.}(2011)Buck, Iwicki, Rossnagel, and
  Kipp]{BuckPRB2011}
Buck,~J.; Iwicki,~J.; Rossnagel,~K.; Kipp,~L. Surface photovoltage effect at
  the $p$-WSe${}_{2}$:Rb surface: Photoemission experiment and numerical model.
  \emph{Phys. Rev. B} \textbf{2011}, \emph{83}, 075312\relax
\mciteBstWouldAddEndPuncttrue
\mciteSetBstMidEndSepPunct{\mcitedefaultmidpunct}
{\mcitedefaultendpunct}{\mcitedefaultseppunct}\relax
\EndOfBibitem
\bibitem[Rossi \latin{et~al.}(2016)Rossi, Büch, Rienzo, Miseikis, Convertino,
  Al-Temimy, Voliani, Gemmi, Piazza, and Coletti]{Rossi2DMater2016}
Rossi,~A.; Büch,~H.; Rienzo,~C.~D.; Miseikis,~V.; Convertino,~D.;
  Al-Temimy,~A.; Voliani,~V.; Gemmi,~M.; Piazza,~V.; Coletti,~C. Scalable
  synthesis of {WS}2 on graphene and h-{BN}: an all-2D platform for
  light-matter transduction. \emph{2D Materials} \textbf{2016}, \emph{3},
  031013\relax
\mciteBstWouldAddEndPuncttrue
\mciteSetBstMidEndSepPunct{\mcitedefaultmidpunct}
{\mcitedefaultendpunct}{\mcitedefaultseppunct}\relax
\EndOfBibitem
\bibitem[Poletto and Frassetto(2010)Poletto, and
  Frassetto]{PolettoApplOptics2010}
Poletto,~L.; Frassetto,~F. Time-preserving grating monochromators for ultrafast
  extreme-ultraviolet pulses. \emph{Appl. Opt.} \textbf{2010}, \emph{49},
  5465--5473\relax
\mciteBstWouldAddEndPuncttrue
\mciteSetBstMidEndSepPunct{\mcitedefaultmidpunct}
{\mcitedefaultendpunct}{\mcitedefaultseppunct}\relax
\EndOfBibitem
\bibitem[Sentef \latin{et~al.}(2014)Sentef, Claassen, Kemper, Moritz, Oka,
  Freericks, and Devereaux]{SentefNatCommun2014}
Sentef,~M.; Claassen,~M.; Kemper,~A.; Moritz,~B.; Oka,~T.; Freericks,~J.;
  Devereaux,~T. Theory of Floquet band formation and local pseudospin textures
  in pump-probe photoemission of graphene. \emph{Nature Communications}
  \textbf{2014}, \emph{6}, 7047\relax
\mciteBstWouldAddEndPuncttrue
\mciteSetBstMidEndSepPunct{\mcitedefaultmidpunct}
{\mcitedefaultendpunct}{\mcitedefaultseppunct}\relax
\EndOfBibitem
\bibitem[Patrick and Choyke(1970)Patrick, and Choyke]{PatrickPhysRevB1970}
Patrick,~L.; Choyke,~W.~J. Static Dielectric Constant of SiC. \emph{Phys. Rev.
  B} \textbf{1970}, \emph{2}, 2255--2256\relax
\mciteBstWouldAddEndPuncttrue
\mciteSetBstMidEndSepPunct{\mcitedefaultmidpunct}
{\mcitedefaultendpunct}{\mcitedefaultseppunct}\relax
\EndOfBibitem
\bibitem[Laturia \latin{et~al.}(2018)Laturia, Van~de Put, and
  Vandenberghe]{Laturia2DMater2018}
Laturia,~A.; Van~de Put,~M.~L.; Vandenberghe,~W.~G. Dielectric properties of
  hexagonal boron nitride and transition metal dichalcogenides: from monolayer
  to bulk. \emph{npj 2D Materials and Application} \textbf{2018}, \emph{2},
  6\relax
\mciteBstWouldAddEndPuncttrue
\mciteSetBstMidEndSepPunct{\mcitedefaultmidpunct}
{\mcitedefaultendpunct}{\mcitedefaultseppunct}\relax
\EndOfBibitem
\bibitem[Walter \latin{et~al.}(2011)Walter, Bostwick, Jeon, Speck, Ostler,
  Seyller, Moreschini, Chang, Polini, Asgari, MacDonald, Horn, and
  Rotenberg]{WalterPRB2011}
Walter,~A.~L.; Bostwick,~A.; Jeon,~K.-J.; Speck,~F.; Ostler,~M.; Seyller,~T.;
  Moreschini,~L.; Chang,~Y.~J.; Polini,~M.; Asgari,~R.; MacDonald,~A.~H.;
  Horn,~K.; Rotenberg,~E. Effective screening and the plasmaron bands in
  graphene. \emph{Phys. Rev. B} \textbf{2011}, \emph{84}, 085410\relax
\mciteBstWouldAddEndPuncttrue
\mciteSetBstMidEndSepPunct{\mcitedefaultmidpunct}
{\mcitedefaultendpunct}{\mcitedefaultseppunct}\relax
\EndOfBibitem
\bibitem[Hwang \latin{et~al.}(2012)Hwang, Siegel, Mo, Regan, Ismach, Zhang,
  Zettl, and Lanzara]{HwangSciRep2012}
Hwang,~C.; Siegel,~D.~A.; Mo,~S.-K.; Regan,~W.; Ismach,~A.; Zhang,~Y.;
  Zettl,~A.; Lanzara,~A. Fermi velocity engineering in graphene by substrate
  modification. \emph{Scientific Reports} \textbf{2012}, \emph{2}, 590\relax
\mciteBstWouldAddEndPuncttrue
\mciteSetBstMidEndSepPunct{\mcitedefaultmidpunct}
{\mcitedefaultendpunct}{\mcitedefaultseppunct}\relax
\EndOfBibitem
\bibitem[Tancogne-Dejean \latin{et~al.}(2020)Tancogne-Dejean, Oliveira,
  Andrade, Appel, Borca, Le~Breton, Buchholz, Castro, Corni, Correa,
  De~Giovannini, Delgado, Eich, Flick, Gil, Gomez, Helbig, H{\"u}bener,
  Jest{\"a}dt, Jornet-Somoza, Larsen, Lebedeva, L{\"u}ders, Marques, Ohlmann,
  Pipolo, Rampp, Rozzi, Strubbe, Sato, Sch{\"a}fer, Theophilou, Welden, and
  Rubio]{TancogneJChemPhys2020}
Tancogne-Dejean,~N. \latin{et~al.}  Octopus, a computational framework for
  exploring light-driven phenomena and quantum dynamics in extended and finite
  systems. \emph{The Journal of Chemical Physics} \textbf{2020}, \emph{152},
  124119\relax
\mciteBstWouldAddEndPuncttrue
\mciteSetBstMidEndSepPunct{\mcitedefaultmidpunct}
{\mcitedefaultendpunct}{\mcitedefaultseppunct}\relax
\EndOfBibitem
\bibitem[Kohn and Sham(1965)Kohn, and Sham]{KohnPhysRev1965}
Kohn,~W.; Sham,~L.~J. Self-Consistent Equations Including Exchange and
  Correlation Effects. \emph{Phys. Rev.} \textbf{1965}, \emph{140},
  A1133--A1138\relax
\mciteBstWouldAddEndPuncttrue
\mciteSetBstMidEndSepPunct{\mcitedefaultmidpunct}
{\mcitedefaultendpunct}{\mcitedefaultseppunct}\relax
\EndOfBibitem
\bibitem[Runge and Gross(1984)Runge, and Gross]{RungePRL1984}
Runge,~E.; Gross,~E. K.~U. Density-Functional Theory for Time-Dependent
  Systems. \emph{Phys. Rev. Lett.} \textbf{1984}, \emph{52}, 997--1000\relax
\mciteBstWouldAddEndPuncttrue
\mciteSetBstMidEndSepPunct{\mcitedefaultmidpunct}
{\mcitedefaultendpunct}{\mcitedefaultseppunct}\relax
\EndOfBibitem
\bibitem[De~Giovannini \latin{et~al.}(2017)De~Giovannini, H\"ubener, and
  Rubio]{DeGiovanniniJChemTheoryComput2017}
De~Giovannini,~U.; H\"ubener,~H.; Rubio,~A. A First-Principles Time-Dependent
  Density Functional Theory Framework for Spin and Time-Resolved
  Angular-Resolved Photoelectron Spectroscopy in Periodic Systems. \emph{J.
  Chem. Theory Comput.} \textbf{2017}, \emph{13}, 265\relax
\mciteBstWouldAddEndPuncttrue
\mciteSetBstMidEndSepPunct{\mcitedefaultmidpunct}
{\mcitedefaultendpunct}{\mcitedefaultseppunct}\relax
\EndOfBibitem
\bibitem[Hartwigsen \latin{et~al.}(1998)Hartwigsen, Goedecker, and
  Hutter]{HartwigsenPRB1998}
Hartwigsen,~C.; Goedecker,~S.; Hutter,~J. Relativistic separable dual-space
  Gaussian pseudopotentials from H to Rn. \emph{Phys. Rev. B} \textbf{1998},
  \emph{58}, 3641--3662\relax
\mciteBstWouldAddEndPuncttrue
\mciteSetBstMidEndSepPunct{\mcitedefaultmidpunct}
{\mcitedefaultendpunct}{\mcitedefaultseppunct}\relax
\EndOfBibitem
\bibitem[{De Giovannini, Umberto} \latin{et~al.}(2015){De Giovannini, Umberto},
  {Larsen, Ask Hjorth}, and {Rubio, Angel}]{DeGiovanniniEurPhysJB2015}
{De Giovannini, Umberto},; {Larsen, Ask Hjorth},; {Rubio, Angel}, Modeling
  electron dynamics coupled to continuum states in finite volumes with
  absorbing boundaries. \emph{Eur. Phys. J. B} \textbf{2015}, \emph{88},
  56\relax
\mciteBstWouldAddEndPuncttrue
\mciteSetBstMidEndSepPunct{\mcitedefaultmidpunct}
{\mcitedefaultendpunct}{\mcitedefaultseppunct}\relax
\EndOfBibitem
\bibitem[Madsen(2005)]{MadsenAmJPhys2005SI}
Madsen,~L.~B. Strong-field approximation in laser-assisted dynamics.
  \emph{American Journal of Physics} \textbf{2005}, \emph{73}, 57--62\relax
\mciteBstWouldAddEndPuncttrue
\mciteSetBstMidEndSepPunct{\mcitedefaultmidpunct}
{\mcitedefaultendpunct}{\mcitedefaultseppunct}\relax
\EndOfBibitem
\bibitem[Park(2014)]{ParkPhysRevA2014SI}
Park,~S.~T. Interference in Floquet-Volkov transitions. \emph{Phys. Rev. A}
  \textbf{2014}, \emph{90}, 013420\relax
\mciteBstWouldAddEndPuncttrue
\mciteSetBstMidEndSepPunct{\mcitedefaultmidpunct}
{\mcitedefaultendpunct}{\mcitedefaultseppunct}\relax
\EndOfBibitem
\end{mcitethebibliography}

\end{document}